\documentclass[]{interact}

\usepackage{epstopdf}
\usepackage[caption=false]{subfig}

\usepackage[numbers,sort&compress]{natbib}
\bibpunct[, ]{[}{]}{,}{n}{,}{,}
\renewcommand\bibfont{\fontsize{10}{12}\selectfont}

\newcommand{\G}{\mathcal{G}}
\newcommand{\ee}{\mathrm{e}}
\newcommand{\dd}{\mathrm{d}}

\theoremstyle{plain}

\theoremstyle{definition}

\theoremstyle{remark}

\begin{document}


\title{On the two-reactant one-step activation-energy asymptotics for steady, adiabatic, planar flames with Lewis numbers of unity}

\author{
\name{Prabakaran Rajamanickam\thanks{Email: prajaman@ucsd.edu} }
\affil{Department of Mechanical and Aerospace Engineering, University of California San Diego, La Jolla, CA 92093--0411, USA}
}

\maketitle

\begin{abstract}
Aspects of predictions of activation-energy asymptotics concerning the dependence of the burning velocity on the equivalence ratio are examined here through both asymptotic analyses and numerical computation. In typical hydrocarbon-air flames, the burning velocity achieves its maximum value for fuel-rich mixture, the cause being generally attributed to the effects of detailed chemical kinetics and unequal diffusivities of the reactants. The present results demonstrate the possibility of this attribute of the burning velocity occurring even when these two effects are absent. This is accomplished by parametrically studying the burning-velocity formula valid for all equivalence ratios under the conditions specified in the title of this article, with special attention paid to implications for hydrocarbon-air flames.
\end{abstract}

\begin{keywords}
burning velocity; planar flames; activation-energy asymptotics; two-reactant flames; equivalence ratio
\end{keywords}

\section{Introduction}

Concepts of activation-energy asymptotics (AEA) have played important roles in the description of premixed laminar flame structures ever since the work of Zel'dovich and Frank-Kamenetskii~\cite{Zeldovich1938}. Resulting asymptotic formulas for burning velocities of two-reactant flames~\cite{Forman1985}, when plotted as functions of the equivalence ratio, possess attributes that depend on how individual factors in the formulas are varied. Study of the dependence of the burning velocity on the equivalence ratio was initiated by Clarke~\cite{Clarke1975}, for reactants with unity Lewis number. While he anticipated that the burning velocity would reach a maximum for fuel-rich mixtures depending on how the mixture is formed, he did not quantify that idea. It was Sen and Ludford, who carried out the analysis, in a series of publications~\cite{Ludford1979,sen19813,ludford1981,ludford19812,ludford1982}, emphasizing mainly Lewis-number effects and product-dissociation effects, addressing an open question of the late 1970's, namely the extent to which the observed fuel-rich location of the burning-velocity maximum could be attributed to Lewis-number effects rather than to the detailed chemistry. They specifically identified two of the various conditions (to be mentioned later) under which the equivalence ratio can be varied, a constant fraction of inert (case I) and a constant ratio of inert fraction to oxidizer fraction (case II). Most of the experimental burning-velocity measurements that have been reported are for fuel-air mixtures, which correspond to their case II, and which is the condition to be discussed here.

The works of Sen and Ludford emphasized near-stoichiometric conditions, based on the assumption that the peak burning-velocity occurs for slightly fuel-rich conditions, as has been summarized by Bechtold and Matalon~\cite{Matalon1999}. Although this analysis was completed nearly forty years ago, there has been no more recent discussion of their considerations for equivalence ratios not close to unity. It is the purpose of this paper to address all equivalence ratios, without making any reference to Lewis-number effects. For the example of methane-air mixtures, application of leading-order AEA and numerical integration will be shown in this article to demonstrate that, the rich shift predicted by AEA and numerics is large, beyond the range of accuracy of near-stoichiometric AEA, lying instead in the range of the analysis of Clarke, as extended by Mitani~\cite{Mitani1980} and Rogg~\cite{rogg1986}. Prospects for accurate use of AEA for other hydrocarbon-air mixtures also will be considered.

\section{Formulation and asymptotic solution}

Although complicating factors such as variable properties and Stefan-Maxwell transport have been included in previous work~\cite{williams1987}, the points to be addressed here may be based on simpler formulations~\cite{Mitani1980,rogg1986,williams1991}, for one-step Arrhenius chemistry with arbitrary reaction orders $m$ and $n$ with respect to the fuel and oxidizer, respectively. In addition, all Lewis numbers will be set equal to unity, thereby purposely ruling out influences of differential diffusion. Most of the discussion will pertain to $m=n=1$, the values assumed in the work of Clarke~\cite{Clarke1975} and of Sen and Ludford~\cite{Ludford1979}. The gas density $\rho$ and the thermal diffusivity $D_T$ are both constant in the formulation and in the numerical integrations to be reported.

Under the given approximations, a temperature-explicit formulation applies. With $T_o$ and $T_\infty$ denoting the fresh-mixture and burnt-gas temperatures, the normalized dependent variable for the temperature $T$ is $\tau=(T-T_o)/(T_\infty-T_o)$, and the parameter $\alpha=(T_\infty-T_o)/T_\infty$ measures the heat release. In terms of the laminar burning velocity $S_L$, the characteristic length $D_T/S_L$ is introduced to define the nondimensional spatial coordinate $x$. The symbol $\phi$ will be employed for the conventional fuel-air equivalence ratio, so that $0< \phi<\infty$. In terms of the activation energy $E$ and the universal gas constant $R$, the Zel'dovich number, $\beta=\alpha E/(RT_\infty)$, is the large parameter of expansion. Given an appropriate characteristic reciprocal-time pre-factor constant for the reaction rate, $B$, the burning-rate eigenvalue is $\Lambda=(B D_T/S_L^2)\ee^{-E/(RT_\infty)}$. The differential equation to be solved, for instance for a lean mixture, then becomes
\begin{equation}
    \frac{\dd^2\tau}{\dd x^2} = \frac{\dd\tau}{\dd x} - \Lambda (1-\tau)^m (1-\phi\tau)^n \exp\left[- \frac{\beta(1-\tau)}{1-\alpha( 1-\tau)}\right],
\end{equation}
subject to $\tau$ approaching zero as $x$ approaches $-\infty$ and $\tau$ approaching unity as $x$ approaches $+\infty$. 

As is well known, in the limit of $\beta$ approaching infinity, there is an upstream convective-diffusive zone in which $\tau$ is proportional to $\ee^x$, followed by an inner zone, with thickness of order $x/\beta$, that is reactive-diffusive at leading order and within which the order-unity dependent variable $y=\beta(1-\tau)$ must match the convective-diffusive solution as that variable approaches infinity. The problem is independent of $\alpha$ at leading order, when it is of order unity or smaller, and the equation depends on the scaling of $\phi$, the most general choice for fuel-lean or stoichiometric mixtures being that $\gamma_l = \beta(1-\phi)/\phi$ is a parameter of order unity. With this selection, matching at leading order produces
\begin{equation}
    \frac{\beta^{m+n+1}}{2\Lambda \phi^n} = \int_0^\infty t^m(t+\gamma_l)^n \ee^{-t} \dd t \equiv \G(m,n,\gamma_l),
\end{equation}
a result that in fact is also correct when the parameter $\gamma_l$ is large or small in the expansion parameter $\beta$~\cite{Mitani1980}. The function $\G(m,n,\gamma_l)$ is a confluent hypergeometric function, expressible in the form $\G(m,n,\gamma_l)=\gamma_l^{m+n+1}\Gamma(m+1)U(m+1,m+n+2,\gamma_l)$, where $\Gamma$ is the gamma function and $U$ is the Kummer's function of the second kind, and it reduces to  $\Gamma(m+n+1)$ at the stoichiometric condition $\gamma_l=0$, while approaching $\gamma_l^n\Gamma(m+1)$ as $\gamma_l$ approaches infinity. The corresponding result for fuel-rich mixture turns out to be
\begin{equation}
    \frac{\beta^{m+n+1}}{2\Lambda \phi^{1-m}} = \G(n,m,\gamma_r),
\end{equation}
where $\gamma_r=\beta(\phi-1)$.

\section{Variations with equivalence ratio}

 Although some one-step empirical correlations, especially, for autoignition times~\cite{colket2001}, but also occasionally for burning velocities~\cite{westbrook1981}, exhibit negative reaction orders for the fuel, for the great majority of fuels, as well as in studies directed towards revealing qualitative attributes of flame propagation, both $m$ and $n$ are positive. Under these usual conditions, $\G$ achieves a minimum value at $\phi=1$, increasing monotonically in moving away from stoichiometry. The fact that $\G$ usually does not exhibit a maximum value at $\phi=1$ affords the possibility of predicted burning velocities achieving maximum values at conditions far from stoichiometric. The specific form of the function $S_L(\phi)$ for given values of $m$ and $n$ depends on the variations with $\phi$ that are selected for other parameters, such as $B$ and $T_\infty$. The reciprocal time $B$, for example, is proportional to the product of two factors, one being the initial concentration of the oxidizer raised to the power $n$ and the other the initial concentration of the fuel raised to the power $m-1$; at least one of these two factors must be changed to vary $\phi$. In addition, the variation of $T_\infty$ with $\phi$ depends on the specific set of experiments to be addressed. 

The adiabatic flame temperature $T_\infty$ may be held fixed as $\phi$ is changed - a selection often made in counterflow flame experiments to remove the large effect of temperature variations on the chemical kinetics~\cite{seshadri2007}. When that is done, the Arrhenius factor does not influence the function $S_L(\phi)$, but achieving a constant value of $T_\infty$ necessitates decreasing the dilution of the mixture in moving away from the stoichiometric condition $\phi=1$, for typical experiments in which the initial temperature $T_o$ remains constant. There often is interest in varying the stoichiometry at fixed dilution, in which case the influence of the Arrhenius factor on $S_L$ can be dominant, producing a maximum of the predicted burning velocity very close to $\phi=1$ when the activation energy $E$ is large. When realistic values of $E$ and of other parameters are employed in the formula, at constant dilution the maximum of $S_L(\phi)$ often occurs away from stoichiometric conditions, which can be advantageous in fitting burning-velocity data for real flames that achieve maxima at fuel-rich conditions. 

Graphical presentations of computed laminar burning velocities serve to illustrate these results and to test the accuracies of the predictions of the asymptotic formulas. This is done here for a situation in which $\phi$ is varied by isothermal mixing of a fuel stream with an oxidizer stream, both streams being at the same temperature $T_o$. The variation of the adiabatic flame temperature $T_\infty$ with $\phi$ is chosen to correspond to a constant heat capacity for the mixture, thereby determining the variation of $\beta$ with the equivalence ratio. If the mixture is formed by combining diluted fuel and oxidizer streams, then the predicted variations of burning velocities depend on a stoichiometry parameter, the ratio of the mass of the oxygen required to burn the fuel in the fuel stream completely to the actual mass of the oxygen in the oxidizer stream, which will be denoted by $S$, resulting in,
\begin{equation}
    \frac{T_\infty}{T_{\infty,s}} = 1-\alpha_s + \alpha_s \frac{S+1}{S+\phi}\begin{cases}
    \phi ,& \text{for}\ \phi\leq 1\\
                                    1, & \text{for}\ \phi\geq 1,
                                    \end{cases} \label{Tinf}
\end{equation}
 where the subscript $s$ identifies values evaluated at the stoichiometric condition, $\phi=1$. For case I of Sen and Ludford~\cite{Ludford1979}, $S=\nu$, the stoichiometric mass ratio and for case II, $S=\nu(1+b)$, where $b$ is the inert to oxidizer mass ratio; the curves to be shown here correspond to case II. The stoichiometry parameter $S$ defined here will become the natural choice for non-uniform reactant mixtures, such as in premixed wings of the triple flames, upon which the study is motivated. Through its relationship to the adiabatic flame temperature $T_\infty(\phi)$, the variation of the Zel'dovich number and the heat-release parameter can be found from,
\begin{equation}
    \frac{\beta}{\beta_s} = \left(\frac{T_{\infty,s}}{T_\infty}\right)^2\begin{cases}
                                    \phi(S+1)/(S+\phi),& \text{for}\ \phi\leq 1\\
                                    (S+1)/(S+\phi), & \text{for}\ \phi\geq 1,
                                    \end{cases}
\end{equation} 
and 
\begin{equation}
    \frac{\alpha}{\alpha_s} = \frac{\beta}{\beta_s}\frac{T_\infty}{T_{\infty,s}},
\end{equation}
at constant $E$. The reciprocal-time pre-exponential factor defined before becomes
\begin{equation}
    \frac{B}{B_s} = \phi^{m-1} \left(\frac{S+1}{{S+\phi}}\right)^{m+n-1}.
\end{equation}

To increase the generality of the results by avoiding the necessity of selecting particular values for other properties, such as $\rho$ and $D_T$, the figures will show the ratio of the calculated burning velocity to the value of the burning velocity obtained from the (leading-order) asymptotic formula at the stoichiometric point ($\phi=1$), plotted in terms of the equivalence ratio $\phi$. In this scale, the leading-order asymptotic expression for the burning velocity is given by
\begin{subequations} \label{SL}
\begin{equation}
     \phi\leq 1:\  \frac{S_L}{S_{L\infty,s}} =  \left\{\left(\frac{\phi(S+1)}{S+\phi}\right)^{m+n-1} \left(\frac{\beta_s}{\beta}\right)^{m+n+1}   \frac{\G(m,n,\gamma_l)}{\Gamma(m+n+1)} \ \ee^{\beta_s/\alpha_s-\beta/\alpha}\right\}^{1/2}, \label{SL1}
\end{equation}
\begin{equation}
    \phi \geq 1:\  \frac{S_L}{S_{L\infty,s}} =  \left\{\left(\frac{S+1}{S+\phi}\right)^{m+n-1} \left(\frac{\beta_s}{\beta}\right)^{m+n+1}  \frac{\G(n,m,\gamma_r)}{\Gamma(m+n+1)} \ \ee^{\beta_s/\alpha_s-\beta/\alpha}\right\}^{1/2} ,   \label{SL2}            
\end{equation}
\end{subequations}
where the first term in each of the foregoing expressions raised to the power $m+n-1$ is the ratio of upstream concentration of deficient reactant to its stoichiometric value. In the near-stoichiometric limit, only the last two factors in these expressions vary with $\phi$ at leading order, as noted by Sen and Ludford in their analysis~\cite{ludford1982}. Farther away from stoichiometric conditions, however, these terms vary at leading order, and there are other relevant variations, such as $B(\phi)$ and $\beta(\phi)$ that need to be taken into account. No previous publications have shown results which do that.

\section{Representative results}

\begin{figure}
\centering
\includegraphics[scale=0.6]{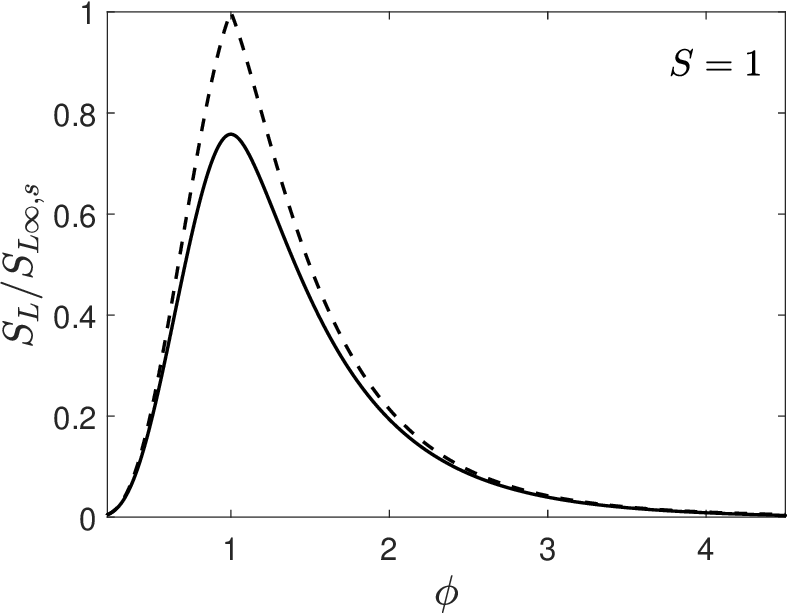}
\caption{Numerical (solid curve) and asymptotic (dashed curve) result for $S=1$ with $\beta_s=8$, $\alpha_s=0.85$ and $(m,n)=(1,1)$.} \label{fig:S1}
\end{figure}

Figure~\ref{fig:S1} compares the leading-order asymptotic prediction (dashed curve) with the result of the numerical integration (solid curve), for the representative values $\beta_s=8$ of the Zel'dovich number of the stoichiometric mixture and $\alpha_s=0.85$ of the heat-release parameter, in the symmetric case $S=1$. Since the Zel'dovich number increases in moving away from stoichiometry, this is its minimum value, whence the asymptotic formula should be increasingly accurate as the departure from $\phi=1$ increases. The figure indicates that expectation to be true and shows that the formula overpredicts the burning-velocity by nearly 30\% at stoichiometric conditions. There is a discontinuity in the slope of the AEA curve of the formulas given above at $\phi=1$ that arises from plotting only the leading-order solution and that can be removed by including suitable terms of order $\beta^{-1}$.

\begin{figure}
\centering
\includegraphics[scale=0.6]{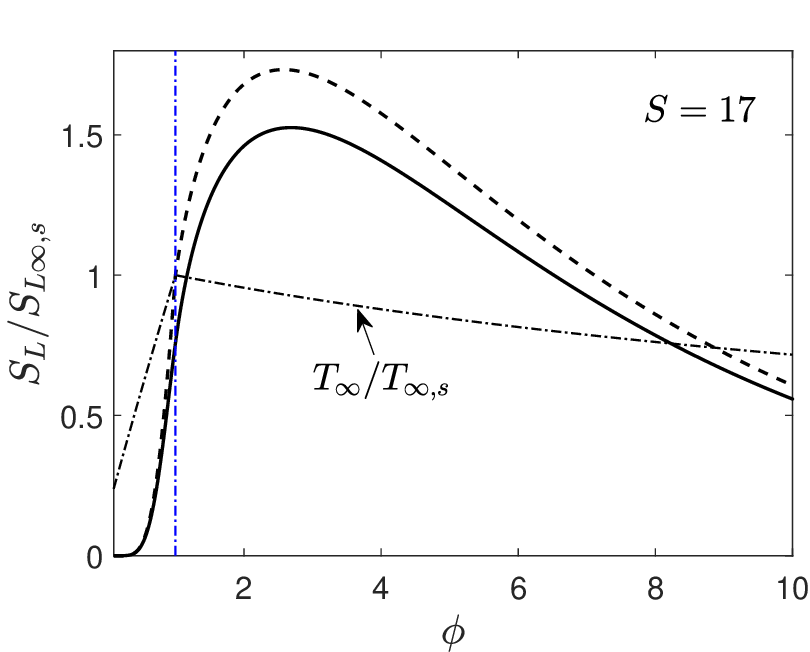}
\caption{Numerical (solid curve) and asymptotic (dashed curve) result for $S=17$ with $\beta_s=8$, $\alpha_s=0.85$ and $(m,n)=(1,1)$.} \label{fig:S17}
\end{figure}

Figure~\ref{fig:S17} shows similar results for $S=17$, the value applicable when the fuel stream is pure methane and the oxidizer stream is air. The values of $\beta_s$ and $\alpha_s$ have been selected to correspond to reasonable flame temperatures and burning-velocity variations. The results seen here, which are much more representative for the combustion of hydrocarbon fuels (and many others) in air, are quite different from those in Fig.~\ref{fig:S1}. The numerical result remains roughly 30\% below the asymptotic prediction at the stoichiometric point, $\phi=1$. This figure illustrates clearly the facts that, not only the result of the numerical integration, but the prediction of the asymptotic formula as well, can give burning velocities that are larger than those for stoichiometric conditions by a significant amount - the differences being of order unity. For the asymptotic prediction, this behaviour is due entirely to the variation of the function $\G$, the variation of the Arrhenius factor with $T_\infty$ at the fixed value of $E$ opposing this effect but not strong enough to overcome it in rich flames, as may be seen from the $T_\infty$ curve in Fig.~\ref{fig:S17}. The dilution does decrease with $\phi$ in this mixing process when $S>1$, but that decrease is not great enough to produce a decrease in $T_\infty$. This figure also illustrates that, with $\beta_s=8$ and $m=n=1$, the equivalence ratio at which the laminar burning velocity is maximum is in close agreement for asymptotic and numerical results, but it exceeds the value that typically would be obtained using the correct detailed chemistry for methane, and it occurs at a value of $\phi$ for which predictions of near-stoichiometric AEA would be highly inaccurate. The predictions shown here are found to differ by approximately 25\% from the near-stoichiometric expansion of~\eqref{SL}, a result that is not plotted here. These far-from-stoichiometric results are not addressed in the earlier publications, such as those of Sen and Ludford.

\begin{figure}
\centering
\includegraphics[scale=0.5]{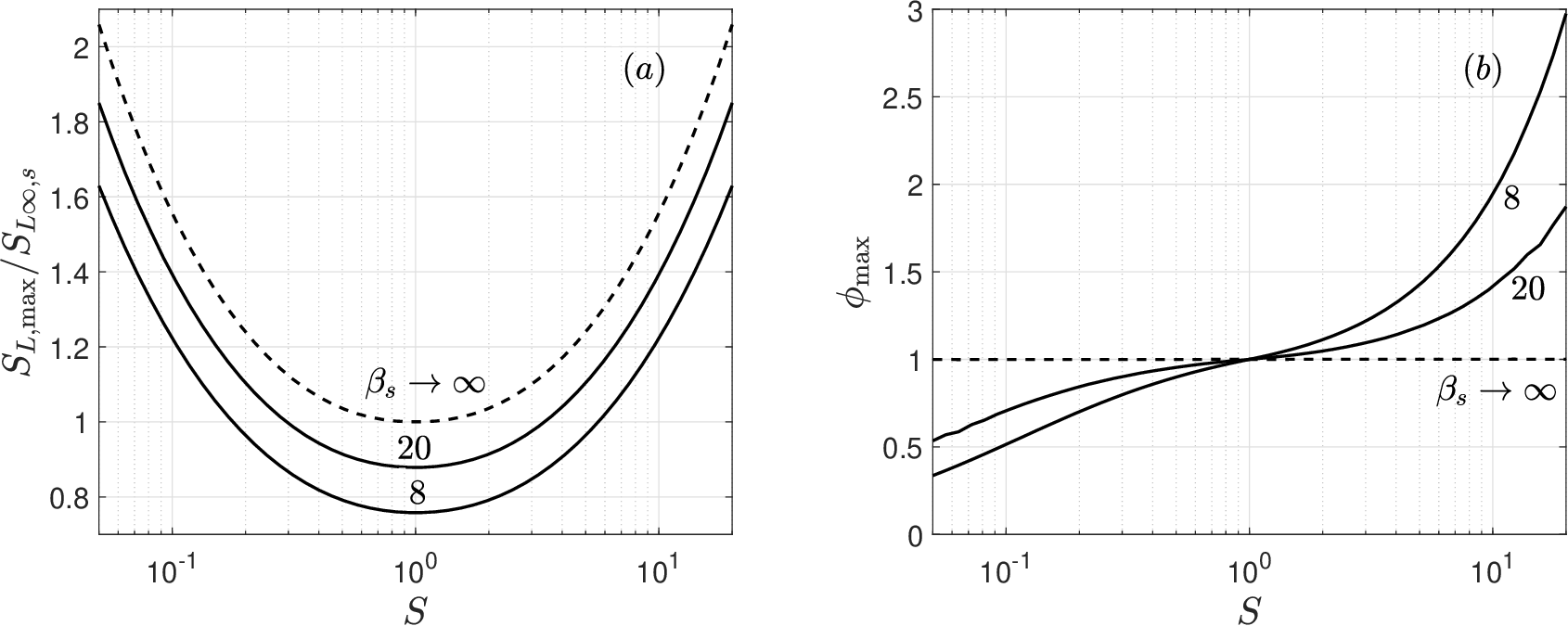}
\caption{Maximum burning-velocity and its location as a function of stoichiometric ratio, $S$ with $\alpha_s=0.85$ and $(m,n)=(1,1)$.} \label{fig:max}
\end{figure}

\section{Discussions}

 A general observation of this study is that influences of the Arrhenius factor decrease compared with influences of $\G$ as departures from $S=1$ increase. For $m=n$, the magnitudes of departures are the same at the same value of $|\ln S|$, whether $\ln S$ is positive or negative as shown in Fig.~\ref{fig:max}(a), but this symmetry is lost for $m\neq n$. Figure~\ref{fig:max}(b) shows how the equivalence ratio at which the maximum burning velocity is achieved varies increasingly strongly with $S$ as $\beta_s$ decreases. The large increase in the burning velocity in rich flames, above its value at stoichiometric condition, shown in Fig.~\ref{fig:S17}, demonstrates how poor this one-step Arrhenius chemistry approximates methane-air flames. In this case the burning-velocity maximum occurs well beyond the range of accuracy of a near-stoichiometric expansion and at much more fuel-rich condition than found experimentally. Although this reaction-rate approximation is poor for methane-air, it may be better for other hydrocarbon-air mixtures, such as ethylene-air, for which the burning-velocity maximum occurs at higher equivalence ratios. For ethylene-air flames, Lewis numbers are close enough to unity for the assumptions of the present formulation to apply, but for flames of propane and higher hydrocarbons, effects of differential diffusions, excluded here, might be expected to become increasingly important, although Fig. 7.7.4, on page 277 of the textbook by Law~\cite{law2010}, showing essentially identical burning-velocity curves for higher normal alkanes when plotted against the equivalence ratio, suggests that this effect may not be noticeably large. 
 
 \section{Conclusions}
 
 The Arrhenius factor is not always dominant in AEA predictions, in that the factors may produce off-stoichiometric burning-rate maxima even without differential diffusion. Formulas of AEA may provide reasonable fits to burning-velocity data for some hydrocarbon-air mixtures, such as ethylene-air systems, but such results are inaccurate for methane-air mixtures. In addition, reaction orders $m$ and $n$ can be adjusted to fit to burning-velocities of different hydrocarbon-air mixtures, although this is not addressed here.

\section*{Acknowledgements}

Professor F.A. Williams, who advised the writing of this communication, made significant contributions towards the investigation. Professor A.L. S\'anchez took part in initiating the problem, for which the author is grateful. The author would also like to thank two anonymous reviewers and the editor for a number of suggestions that led to significant improvements in this article.

\section*{Disclosure statement}

No potential conflict of interest was reported by the author.

\bibliographystyle{tfq}
\bibliography{interacttfqsample}

\begin{thebibliography}{10}
\newcommand{\printfirst}[2]{#1}
\newcommand{\switchargs}[2]{#2#1}
\providecommand{\url}[1]{\normalfont{#1}}
\providecommand{\urlprefix}{Available at }

\bibitem{Zeldovich1938}
Y.B. Zel'dovich and D.A. Frank-Kamenetskii, \emph{A theory of thermal
  propagation of flame}, Zh. Fiz. Khim+. 12 (1938), pp. 100--105.

\bibitem{Forman1985}
F.A. Williams, \emph{Combustion Theory}, 2nd Edn., Benjamin/Cummings., 1985,
  pp.154-165.

\bibitem{Clarke1975}
J.F. Clarke, \emph{The pre-mixed flame with large activation energy and
  variable mixture strength: Elementary asymptotic analysis}, Combust. Sci.
  Technol. 10 (1975), pp. 189--194.

\bibitem{Ludford1979}
A.K. Sen and G.S.S. Ludford, \emph{The near-stoichiometric behavior of
  combustible mixtures part i: Diffusion of the reactants}, Combust. Sci.
  Technol. 21 (1979), pp. 15--23.

\bibitem{sen19813}
A.K. Sen and G.S.S. Ludford, \emph{The near-stoichiometric behavior of
  combustible mixtures part ii: Dissociation of the products}, Combust. Sci.
  Technol. 26 (1981), pp. 183--191.

\bibitem{ludford1981}
A.K. Sen and G.S.S. Ludford, \emph{Effects of mass diffusion on the burning
  rate of non-dilute mixtures}, Symp. (Int.) Combust. 18 (1981), pp. 417--424.

\bibitem{ludford19812}
G.S.S. Ludford and A.K. Sen, \emph{Burning rate maximum of a plane premixed
  flame}, Prog. Astronaut. Aeronaut. 76 (1981), pp. 427--436.

\bibitem{ludford1982}
A.K. Sen and G.S.S. Ludford, \emph{Maximum flame temperature and burning rate
  of combustible mixtures}, Symp. (Int.) Combust. 19 (1982), pp. 267--274.

\bibitem{Matalon1999}
J.K. Bechtold and M. Matalon, \emph{Effects of stoichiometry on stretched
  premixed flames}, Combust. Flame 119 (1999), pp. 217--232.

\bibitem{Mitani1980}
T. Mitani, \emph{Propagation velocities of two-reactant flames}, Combust. Sci.
  Technol. 21 (1980), pp. 175--177.

\bibitem{rogg1986}
B. Rogg, \emph{On the accuracy of asymptotic flame speed predictions for
  two-reactant flames}, Combust. Sci. Technol. 45 (1986), pp. 317--329.

\bibitem{williams1987}
H.K. Chelliah and F.A. Williams, \emph{Asymptotic analysis of two-reactant
  flames with variable properties and stefan-maxwell transport}, Combust. Sci.
  Technol. 51 (1987), pp. 129--144.

\bibitem{williams1991}
F.A. Williams, \emph{Overview of asymptotics for methane flames}, in
  \emph{Reduced Kinetic Mechanisms and Asymptotic Approximations for
  Methane-Air Flames}, Springer,  1991, pp. 68--85.

\bibitem{colket2001}
M.B. Colket and L.J. Spadaccini, \emph{Scramjet fuels autoignition study}, J.
  Propul. Power 17 (2001), pp. 315--323.

\bibitem{westbrook1981}
C.K. Westbrook and F.L. Dryer, \emph{Simplified reaction mechanisms for the
  oxidation of hydrocarbon fuels in flames}, Combust. Sci. Technol. 27 (1981),
  pp. 31--43.

\bibitem{seshadri2007}
R. Seiser, S. Humer, K. Seshadri, and E. Pucher, \emph{Experimental
  investigation of methanol and ethanol flames in nonuniform flows}, Proc.
  Combust. Inst. 31 (2007), pp. 1173--1180.

\bibitem{law2010}
C.K. Law, \emph{Combustion Physics}, Cambridge university press, 2010.

\end{thebibliography}

\end{document}